\documentclass{eptcs}

\usepackage{amssymb}
\usepackage{amsmath}
\usepackage{stmaryrd}
\usepackage{xspace}
\usepackage{listings}
\usepackage{breakurl}

\newif\ifsubmit
\submitfalse

\ifsubmit
\newcommand{\nb}[2]{}
\else
\newcommand{\nb}[2]{
    \fbox{\bfseries\sffamily\scriptsize#1}
    {\sf\small$\blacktriangleright$\textit{#2}$\blacktriangleleft$}
   }
\fi

\newcommand{\C}{\mathit{c}} 
\newcommand{\tr}{\mathit{t}} 
\newcommand{\atr}{\mathit{u}} 

\newcommand{\var}{\ensuremath{\mathit{X}\xspace}}
\newcommand{\avar}{\ensuremath{\mathit{Y}\xspace}}

\newcommand{\val}{\ensuremath{\mathit{v}\xspace}}
\newcommand{\num}{\ensuremath{\mathit{i}\xspace}}

\newcommand{\intr}[1]{\llbracket{#1}\rrbracket}
\newtheorem{definition}{Def.}[section]
\newtheorem{theorem}{Theorem}[section]
\newtheorem{lemma}{Lemma}[section]
\newtheorem{corollary}{Corollary}[section]

\newcommand{\zero}{\ensuremath{\mathit{zero}\xspace}}
\newcommand{\scc}{\ensuremath{\mathit{succ}\xspace}}
\newcommand{\prd}{\ensuremath{\mathit{pred}\xspace}}
\newcommand{\zer}{\ensuremath{\mathit{zer}\xspace}}
\newcommand{\nat}{\ensuremath{\mathit{nat}\xspace}}
\newcommand{\pos}{\ensuremath{\mathit{pos}\xspace}}
\newcommand{\evn}{\ensuremath{\mathit{evn}\xspace}}
\newcommand{\odd}{\ensuremath{\mathit{odd}\xspace}}
\newcommand{\add}{\ensuremath{\mathit{add}\xspace}}
\newcommand{\Int}{\mathbb{Z}}
\newcommand{\dv}{\ensuremath{\mathit{d}\xspace}} 
\newcommand{\fun}{\ensuremath{\mathcal{F}\xspace}}
\newcommand{\lemmafun}{\ensuremath{\fun_L}}
\newcommand{\emptyType}[1]{\ensuremath{{#1}\downarrow_{\bot}}}
\newcommand{\notEmptyType}[1]{\ensuremath{{#1}\uparrow_{\bot}}}
\newcommand{\negNotEmptyType}[1]{\ensuremath{{#1}\not{\hspace*{-1.2pt}\uparrow}_{\bot}}}
\newcommand{\proof}{\paragraph{Proof:}}
\newcommand{\qed}{\hspace*{\fill}$\Box$}
\newcommand{\object}{\ensuremath{\mathit{object}}\xspace}

\newcommand{\removed}[1]{}

\newcommand{\Pair}[2]     {({#1},{#2})}
\newcommand{\Rule}[4]{\scriptstyle{\textrm{({#1})}}\displaystyle\frac{#2}{#3}\ #4}

\newcommand{\production}[2]{{#1} & ::= & {#2} & \\ }

\newcommand{\vbar}{\mid}

\newcommand{\m}{\mathit{m}}

\newcommand{\f}{\mathit{f}} 
\newcommand{\g}{\mathit{g}} 

\newcommand{\this}{\mathit{this}}

\newcommand{\methpred}{\mathit{has\_meth}}

\newcommand{\extendspred}{\mathit{extends}}
\newcommand{\invokepred}{\mathit{invoke}}

\newcommand{\fieldaccpred}{\mathit{field\_acc}}

\newcommand{\newpred}{\mathit{new}}

\newcommand{\obj}{\mathit{obj}}

\newcommand{\intType}{\mathit{int}}

\newcommand{\objectType}[2]{\obj\Pair{#1}{#2}}

\newcommand{\union}{\vee}
\newcommand{\unionType}[2]{#1\union #2}

\newcommand{\subtype}{\leq}

\begin{document}
\title{Coinductive subtyping for abstract compilation of object-oriented languages into Horn formulas\thanks{This work has been partially
supported by MIUR DISCO - Distribution, Interaction, Specification, Composition for Object Systems.}
}
\author{Davide Ancona
\institute{DISI, University of Genova\\Italy}
\email{davide@disi.unige.it}
\and
Giovanni Lagorio
\institute{DISI, University of Genova\\Italy}
\email{lagorio@disi.unige.it}
}
\def\titlerunning{Coinductive subtyping for abstract compilation of object-oriented languages into Horn formulas}
\def\authorrunning{D. Ancona, G. Lagorio}

\maketitle

\begin{abstract}
In recent work we have shown how it is possible to
define very precise type systems for object-oriented languages
by abstractly compiling a program into a Horn formula $f$.
Then type inference amounts to resolving a certain goal
w.r.t. the coinductive (that is, the greatest) Herbrand model
of $f$. 

Type systems defined in this way are idealized,
since in the most interesting instantiations both the terms of the coinductive Herbrand universe and 
goal derivations cannot be finitely represented. However, sound and quite expressive approximations 
can be implemented by considering only regular terms and derivations. In doing so, it is essential to introduce
a proper \emph{subtyping} relation formalizing the notion of approximation between types.

In this paper we study a subtyping relation on coinductive terms built on union and object type constructors. 
We define an interpretation of types as set of values induced by a quite intuitive relation of membership of values to types, and 
prove that the definition of subtyping is sound w.r.t. subset inclusion between type interpretations.    
The proof of soundness has allowed us to simplify the notion of contractive derivation and to 
discover that the previously given definition of subtyping did not cover all possible representations
of the empty type.
\end{abstract}

\section{Introduction}\label{sect:intro}

In recent work \cite{AnconaEtAl09} we have defined a framework which allows
precise type analysis of object-oriented programs by 
means of abstract compilation of the program to be analyzed into
a Horn formula (that is, a conjunction of Horn clauses).
Then, type inference corresponds to resolving a certain goal (or query)
w.r.t. the coinductive (that is, the greatest) Herbrand model
of $f$. 

Coinductively defined terms of the Herbrand universe (which correspond to type expressions), 
in conjunction with the union type constructor, provide an abstract representation for arbitrary sets of values,
whereas coinductive SLD resolution \cite{SimonEtAl06,SimonEtAl07} allows type inference of recursive method invocation.  
However, type systems defined in this way are idealized,
since, except for the most simple cases where types are just constants, in the most interesting instantiations both terms and 
goal derivations cannot be finitely represented.

However, sound and quite expressive approximations 
can be implemented by considering only regular types and derivations, that is, infinite terms and trees, respectively, which can be finitely represented. 
In doing so, it is essential to introduce
a proper \emph{subtyping} relation \cite{AnconaLagorio09} formalizing the notion of approximation between types, and a corresponding notion of
\emph{subsumption} at the level of goal derivation. 
In this way, regular types, which correspond to usual recursive types, are simply considered as approximations (that is, supertypes)
of much finer infinite types which have no finite representation.

This novel approach has several advantages:
\begin{itemize}
\item It offers a quite general and highly modular framework for type analysis of object-oriented programs, where quite different
kinds of analysis can be defined without changing the core inference engine based on coinductive SLD resolution empowered by the notions of subtyping and subsumption. 
Every instantiation corresponds to a particular choice of the type constructors, the abstract compilation schema, and the definition of the subtyping relation.
Our previous papers provide several examples corresponding to different instantiations of the same framework \cite{AnconaLagorio09,AnconaLagorio10};
under this point of view, our proposal is an attempt to provide a common framework for reasoning on type analysis of object-oriented programs. Indeed,
the solutions to the problem of type analysis of object-oriented programs which can be found in literature \cite{PalsbergSchwartzbach91,OxhojEtAl92,Ages95a,WangSmith01,WangSmith08,Furr09}
are often rather ad hoc, cannot be easily described in an abstract way, and, for these reasons, cannot be easily compared.

\item Several static analysis techniques for compiler optimization can be easily adopted for enhancing type analysis. For instance, we have shown \cite{AnconaLagorio10} 
that a more precise type analysis can be obtained when abstract compilation 
is performed on programs in Static Single Assignment intermediate form \cite{CytronEtAl91}.

\item It promotes a nice integration between theory and practice, since type inference algorithms are just approximations of an idealized type
system where its derivable type judgments can be expressed as the limits of chains of approximating judgments derivable by the algorithm, where their precision depends on the space and 
time resources available to the implementation. 
\end{itemize}

The definition of a suitable subtyping relation is of paramount importance to obtain
reasonable approximations of our framework, especially in the presence of union types, which have proved to
be quite expressive when coinductive terms are considered. 

For this reason, in this paper we study a subtyping relation on coinductive terms built on union and object type constructors. Since
types may be infinite, the relation is defined coinductively; however, such a definition
is far from being intuitive, because a suitable notion of \emph{contractive} \cite{BrandtHenglein97,BrandtHenglein98} derivation has to be
introduced to avoid unsound derivations. 
The contributions of this paper w.r.t. our previous work are the following:
\begin{itemize}
\item We define an interpretation of types
as set of values induced by a quite intuitive relation of membership of values to types.
\item We prove that the definition of subtyping is sound w.r.t. subset inclusion between type interpretations.
The proof of soundness has allowed us to simplify the notion of contractive derivation for subtyping.
\item We have discovered that the previously given definition of subtyping did not cover all possible representations
of types with an empty interpretation.
Consequently, a new subtyping rule has been added, based on a complete characterization
of empty types; such a characterization allowed us to define an algorithm for checking empty regular types.
\end{itemize}

In Section~\ref{sect-ex} a gentle introduction to the framework is given by means of simple examples.
Subtyping and type interpretation are defined in Section~\ref{sect-subtype}, whereas Section~\ref{sect-sound} is devoted to the proof of soundness.
Section~\ref{sect-complete} deals with empty types, and, finally, Section~\ref{sect-conclu} draws some conclusion.         

\section{Abstract compilation into Horn formulas}\label{sect-ex}
Let us consider the standard encoding of natural numbers with objects,
written in Java-like code where, however, all type annotations have been omitted.
\begin{lstlisting}
class Zero {
    add(n)  { return n; }
}

class Succ {
    pred;
    Succ(n) { this.pred=n; }
    add(n)  { return pred.add(new Succ(n)); }
}
\end{lstlisting}
For simplicity, we just consider method \lstinline{add}; class \lstinline{Succ}
represents all natural numbers greater than zero, that is, all numbers
which are successors of a given natural number, stored in the field \lstinline{pred}.    

In the abstract compilation approach a program, as the one shown above, is translated into a
Horn formula where predicates encode the constructs of the language.
For instance, the predicate $\invokepred$ corresponds to method invocation,
and has four arguments: the target object, the method name, the argument list, and
the returned result. Terms represent either types
(that is, set of values) or names (of classes, methods and fields).
In the instantiation we consider here, types include object types $\objectType{\C}{[\f_1{:}\tr_1,\ldots,\f_n{:}\tr_n]}$, where
$\C$ is the class of the object and $\f_1,\ldots,\f_n$ its fields with their corresponding types
$\tr_1,\ldots,\tr_n$, union types $\unionType{\tr_1}{\tr_2}$, and primitive types as $\intType$. 
In the idealized abstract compilation framework, terms can be also infinite and non regular\footnote{We refer to the author's previous work \cite{AnconaEtAl09,AnconaLagorio09,AnconaLagorio10} for more details.}; a regular term is a term which can be infinite, but can only contain a finite number of subterms or, equivalently, 
can be represented as the solution of a unification problem, that is, a finite set of syntactic
equations of the form $X_i=\tr_i$, where all variables $X_i$ are distinct and terms $\tr_i$ may only contain variables $X_i$ \cite{Courcelle83,SimonEtAl06,SimonEtAl07}. 
For instance, the term $\tr$ s.t. $\tr=\unionType{\intType}{\tr}$ is regular\footnote{The exact meaning of such a term will be explained in the next section.} since it has only two subterms, namely, $\intType$ and itself.

Let us see some examples of regular types, that is, regular terms representing set of values.
$$
\begin{small}
\begin{array}{lll}
\zer&=&\objectType{\zero}{[\ ]}\\
\nat&=&\unionType{\zer}{\objectType{\scc}{[\prd{:}\nat]}} \\
\pos&=&\unionType{\objectType{\scc}{[\prd{:}\zer]}}{\objectType{\scc}{[\prd{:}\pos]}} \\
\evn&=&\unionType{\zer}{\objectType{\scc}{[\prd{:}\objectType{\scc}{[\prd{:}\evn]}]}} \\
\odd&=&\objectType{\scc}{[\prd{:}\zer]}\vee\\
&&\objectType{\scc}{[\prd{:}\objectType{\scc}{[\prd{:}\odd]}]} 
\end{array}
\end{small}
$$
Type $zer$ corresponds to all objects representing zero, while $\nat$ corresponds to all objects representing natural numbers and, similarly, $\pos$, $\evn$ and $\odd$ to all objects
representing positive, even, and odd natural numbers, respectively. An example of non regular types is
given by the infinite sequence $\unionType{\tr_1}{(\unionType{\tr_2}{(\unionType{\ldots}{\tr_n}\ldots)})}$, where the term
$\tr_i$ represents the $i^\mathrm{th}$ prime number.
     
Each method declaration is compiled into a single clause, defining a different case for the predicate $\methpred$, that takes
four arguments: the class where the method is declared, its name, the types of its arguments, including
the special argument $\this$ corresponding to the target object, and the type of the returned value.
Predicate $\methpred$ defines the usual method look-up: $\methpred(\C,\m,[\this,\tr_1,\ldots,\tr_n],\tr)$ succeeds
if look-up of $\m$ from class $\C$ succeeds and returns a method that, when invoked on target object
and arguments $\this,\tr_1,\ldots,\tr_n$, returns values of type $\tr$.  

For instance, the method declarations of the two classes defined above are compiled as follows:
\begin{lstlisting}[language=Prolog]
has_meth(zero,add,[This,N],N).

has_meth(succ,add,[This,N],R) $\leftarrow$ 
    field_acc(This,pred,P),
    new(succ,[N],S),
    invoke(P,add,[S],R).
\end{lstlisting}
Predicates $\fieldaccpred$, $\newpred$ and $\invokepred$ correspond to field access, constructor invocation and method invocation, respectively.
Similarly to what happens for methods, each constructor declaration is also compiled into a clause. For instance, the following clause is generated 
from the constructor of class \lstinline{Succ}:
\begin{lstlisting}[language=Prolog]
new(succ,[N],obj(succ,[pred:N|R])) $\leftarrow$ extends(succ,P),new(P,[],obj(P,R)).
\end{lstlisting}
In this case, since  we know\footnote{The set of all clauses generated from the two class declarations is available in the Appendix.} that 
$\extendspred(\scc,\object)$ and 
$\newpred(\object,[\ ],\objectType{\object}{[\ ]})$ hold, then we can derive 
$\newpred(\scc,[N],\objectType{\scc}{[\prd:N]})$.

Other generated clauses are common to all programs and depend on the semantics of the language or on the meaning of types.
\begin{lstlisting}[language=Prolog]
invoke(T1$\vee$T2,M,A,R1$\vee$R2) $\leftarrow$ invoke(T1,M,A,R1), invoke(T2,M,A,R2). 
invoke(obj(C,R),M,A,Res) $\leftarrow$ has_meth(C,M,[obj(C,R)|A],Res). 
\end{lstlisting}
The first clause specifies the behavior of invoke with union types. The invocation must be correct for both target types
$T_1$ and $T_2$ and the returned type is the union of the returned types $R_1$ and $R_2$.
When the target is an object type $\objectType{C}{R}$, then invocation of $M$ with arguments $A$ is correct 
if look-up of $M$ with first argument $\objectType{C}{R}$, corresponding to $\this$, and rest of arguments $A$ succeeds when starting from class $C$.  

We show now that the goal $\invokepred(\evn,\add,[\odd],R)$ is derivable for $R=\tr$ where $\tr$ is the regular
type s.t. $\tr=\unionType{\odd}{\tr}$. If we take for granted that $\tr$ is equivalent\footnote{The equivalence between the two terms will be clarified in the next section.} 
to $\odd$, then not only we can prove that adding
an even and an odd number always returns an odd number, but we can also infer the thesis (that is, the result is an odd number),
since the query corresponds to just asking which number is returned when adding an even and an odd number. 

We recall that, when considering the coinductive Herbrand model, derivations are allowed to be infinite \cite{SimonEtAl06}. 
Then, since $\evn=\unionType{\zer}{\objectType{\scc}{[\prd{:}\objectType{\scc}{[\prd{:}\evn]}]}}$, by clause 1
for $\invokepred$ we must show that $\invokepred(\zer,\add,[\odd],\odd)$ and
$\invokepred(\objectType{\scc}{[\prd{:}\objectType{\scc}{[\prd{:}\evn]}]},\add,[\odd],\tr)$. \linebreak
The first atom can be derived by applying clause 2 for $\invokepred$, and then the clause for $\methpred$
generated from class \lstinline{Zero}. For the second atom we apply clause 2 for $\invokepred$, and then the clause for $\methpred$
generated from class \lstinline{Succ} and get $\invokepred(\objectType{\scc}{[\prd{:}\evn]},\add,[\objectType{\scc}{[\prd{:}\odd]}],\tr)$.
Then, if we re-apply the same clauses once again, we get $\invokepred(\evn,\add,[\scc^2(\odd)],\tr)$
(where $\scc^2(\odd)$ is just an abbreviation for $\objectType{\scc}{[\prd{:}\objectType{\scc}{[\prd{:}\odd]}]}$)
which is equal to the initial goal, except for the argument type which is $\scc^2(\odd)$ instead of $\odd$. 
It is now clear that we can get an infinite derivation containing all atoms having shape $\invokepred(\evn,\add,[\scc^{2n}(\odd)],\tr)$
for all $n\geq 0$, hence $\invokepred(\evn,\add,[\odd],\tr)$ is derivable.

There are two main problems with the example of derivation given above: it is not regular, hence it cannot be computed, and
we would like to resolve $\invokepred(\evn,\add,[\odd],R)$ for $R=\odd$ rather than for $R=\tr$. To overcome these problems, a subtyping relation has to be introduced
together with a notion of subsumption between atoms. The definition of the subtyping relation is postponed to the next section, however
the intuition suggests that $\scc^2(\odd)\leq\odd$ and $\tr\leq\odd$ should hold.\footnote{More precisely, both directions of the two disequalities hold, since both
pairs of terms are equivalent, but here we are only interested in one specific direction.} Furthermore, the following subsumption relations are expected to hold:
if $\scc^2(\odd)\leq\odd$, then $\invokepred(\evn,\add,[\odd],\tr)$ subsumes $\invokepred(\evn,\add,[\scc^2(\odd)],\tr)$,
that is, subtyping is contravariant w.r.t. method arguments, as usual, and, therefore, if method $\add$ returns $\tr$ when applied to argument $\odd$, then
it returns $\tr$ when applied to any subtype of $\odd$ (in this specific case, $\scc^2(\odd)$).   
On the other hand, subtyping is covariant w.r.t. the returned type, therefore if $\tr\leq\odd$ then $\invokepred(\evn,\add,[\odd],\tr)$ subsumes $\invokepred(\evn,\add,[\odd],\odd)$,
that is, if method $\add$ returns $\tr$ when applied to $\odd$, then it returns all supertypes of $\tr$ as well ($\odd$ in this specific case).

By introducing subtyping and subsumption it is possible to build a regular derivation for $\invokepred(\evn,$ $\add,[\odd],\tr)$, by just  
observing that to prove $\invokepred(\evn,\add,[\odd],\tr)$ we need to prove $\invokepred(\evn,\add,$ $[\scc^2(\odd)],\tr)$ which, in turn, is subsumed
by $\invokepred(\evn,\add,[\odd],\tr)$, hence we can conclude the proof by coinductive hypothesis. Finally, by applying subsumption
once more we can derive $\invokepred(\evn,\add,$ $[\odd],\odd)$ from $\invokepred(\evn,\add,[\odd],\tr)$.
More in practice, this means that coSLD resolution~\cite{SimonEtAl06} can be generalized by taking into account subtyping constraints between terms, besides the usual unification constraints.

\section{Subtyping and type interpretation}\label{sect-subtype}
In this section we formally define subtyping as a syntactic relation between types; then we provide 
an intuitive interpretation of types
as sets of values, to define a semantic counterpart of the subtyping relation. 

\subsection{Definition of subtyping}
The types we consider are all infinite terms coinductively defined as follows:
$$
\begin{array}{rcll}
\production{\tr}{\intType \vbar \objectType{\C}{[\f_1{:}\tr_1,\ldots,\f_n{:}\tr_n]} \vbar \unionType{\tr_1}{\tr_2}} 
\end{array}
$$
An object type $\objectType{\C}{[\f_1{:}\tr_1,\ldots,\f_n{:}\tr_n]}$ specifies the class $\C$ to which the object belongs,
together with the set of available fields with their corresponding
types. The class name is needed for typing method invocations.
We assume that fields in an object type are finite, distinct and that their order is immaterial. Union types $\unionType{\tr_1}{\tr_2}$
have the standard meaning \cite{BarbaneraEtAl95,IgarashiNagira07}. 

The subtyping relation is coinductively defined by the rules in Figure~\ref{fig-subtype}.
Rules are conceived for a purely functional setting
\cite{AnconaLagorio09}, an extension for dealing with imperative
features can be found in another paper \cite{AnconaLagorio10} by the same authors.

\begin{figure*}[t]
\begin{small}
\begin{center}
\begin{math}
\begin{array}{c}
\Rule{int}{}{\intType\leq\intType}{}\qquad
\Rule{$\vee$R1}{\tr\leq\tr_1}{\tr\leq\unionType{\tr_1}{\tr_2}}{}
\qquad 
\Rule{$\vee$R2}{\tr\leq\tr_2}{\tr\leq\unionType{\tr_1}{\tr_2}}{}
\qquad
\Rule{$\vee$L}{\tr_1\leq\tr\quad\tr_2\leq\tr}{\unionType{\tr_1}{\tr_2}\leq\tr}{}
\\[4ex]
\Rule{obj}
{\tr_1\leq\tr'_1,\ldots,\tr_n\leq\tr'_n}
{\objectType{\C}{[\f_1{:}\tr_1,\ldots,\f_n{:}\tr_n,\ldots]}\leq\objectType{\C}{[\f_1{:}\tr'_1,\ldots,\f_n{:}\tr'_n]}}
{}
\\[4ex]
\Rule{distr}
{
\begin{array}{c}
\objectType{\C}{[\f{:}\atr_1,\f_1{:}\tr_1,\ldots,\f_n{:}\tr_n]}\leq\tr\\
\objectType{\C}{[\f{:}\atr_2,\f_1{:}\tr_1,\ldots,\f_n{:}\tr_n]}\leq\tr
\end{array}
}
{\objectType{\C}{[\f{:}\unionType{\atr_1}{\atr_2},\f_1{:}\tr_1,\ldots,\f_n{:}\tr_n]}\leq\tr}
{}
\end{array}
\end{math}
\caption{Rules defining the subtyping relation}\label{fig-subtype}
\end{center}
\end{small}
\end{figure*}

Rules ($\vee$R1), ($\vee$R2) and ($\vee$L) specify subtyping between
union types, and simply state that the union type constructor is the join operator
w.r.t. subtyping. Note also the strong analogy with the left and right logical rules
of the classical Gentzen sequent calculus for the disjunction, when the subtping relation
is replaced with the provability relation.

Rule (obj) corresponds to standard width and depth subtyping between
object types: the type on the left-hand side may have more fields
(represented by the ellipsis at the end), while subtyping
is covariant w.r.t. the fields belonging to both types. 
Note that depth subtyping is allowed since we are considering a purely
functional setting \cite{AnconaLagorio10}. 
Finally, subtyping between object types is
allowed only when they refer to the same class name.

Rule (distr) expresses distributivity of object over union types;
intuitively, object types correspond to Cartesian product which
distributes over union: $A\times(B\cup C)=(A\times B)\cup(A\times C)$.
For instance $\objectType{\C}{[\f{:}\unionType{\tr_1}{\tr_2}]}\cong
\unionType{\objectType{\C}{[\f{:}\tr_1]}}{\objectType{\C}{[\f{:}\tr_2]}}$, where $\atr_1\cong\atr_2$ holds
iff $\atr_1\subtype\atr_2$ and $\atr_2\subtype\atr_1$.
The relation
$\unionType{\objectType{\C}{[\f{:}\tr_1]}}{\objectType{\C}{[\f{:}\tr_2]}}
\leq\objectType{\C}{[\f{:}\unionType{\tr_1}{\tr_2}]}$ can be derived
by applying rules ($\vee$L), (obj), ($\vee$R1) and ($\vee$R2), and by the fact that
$\tr_1\leq\unionType{\tr_1}{\tr_2}$ and 
$\tr_2\leq\unionType{\tr_1}{\tr_2}$ hold
by reflexivity, which is ensured by
rules (int) and (obj).  
Rule (distr) is necessary for deriving the opposite direction of the relation, since by applying rules ($\vee$R1), ($\vee$R2)
and (obj) we end up with $\unionType{\tr_1}{\tr_2}\leq\tr_1$ or $\unionType{\tr_1}{\tr_2}\leq\tr_2$ which in general do not hold.
Finally, note that rule (distr) is applicable only when the object
type on the left-hand side has at least a field associated with a union
type; since order of fields is immaterial, in the rule such a field
appears always in the first position for readability.   

A derivation is a tree where each node is a pair consisting of
a judgment of the shape $\tr_1\leq\tr_2$, and the label of a
rule\footnote{This labeling is necessary for the proof of soundness.},
and where each node, together with its children, corresponds to a valid
instantiation of a rule. For instance, the following tree
$$
\begin{array}{ccccc}
(\intType\leq\intType,\mbox{int}) & &  & & (\intType\leq\intType,\mbox{int}) \\
& \nwarrow & & \nearrow & \\
&&(\unionType{\intType}{\intType}\leq\intType,\mbox{$\vee$L})&&
\end{array}
$$
is a derivation for $\unionType{\intType}{\intType}\leq\intType$.
However, in the rest of the paper we will use the following equivalent
but more intuitive representation for derivations:
$$
\mbox{\scriptsize ($\vee$L)}\dfrac
    {
      \begin{array}{cc}
        \mbox{\scriptsize (int)}\dfrac{}{\intType\leq\intType}  &  \mbox{\scriptsize (int)}\dfrac{}{\intType\leq\intType}
      \end{array}
    }
    {\unionType{\intType}{\intType}\leq\intType}
$$

Since subtyping is defined over infinite types, all rules must be
interpreted coinductively, therefore derivations are allowed to be
infinite.
However, not all infinite derivations can be  considered valid, but only those \emph{contractive} \cite{BrandtHenglein97,BrandtHenglein98} (see the definition below).
To see why we need such a restriction, consider the regular type $\atr$ s.t. $\atr=\unionType{\atr}{\atr}$, and the following infinite derivation containing just applications
of rules ($\vee$R1) and ($\vee$R2):
$$
\dfrac{\dfrac{\vdots}{\intType\leq\atr}}
{\intType\leq\atr}
$$
We reject infinite derivations built applying only rules ($\vee$R1) and ($\vee$R2), since 
they allow unsound judgments, as $\intType\leq\atr$ derived
above. As it will be shown in Section~\ref{sect-member},
$\atr$ corresponds to the empty type, that is, to the bottom element
$\bot$ w.r.t. the subtyping relation; indeed, for any type $\tr$ there
exists a contractive derivation for $\bot\leq\tr$ obtained  
by applying rule ($\vee$L) infinite times.

Before giving the formal definition of contractive derivation, let us consider another example: if $\bot$ is again the regular type s.t. $\bot=\unionType{\bot}{\bot}$, then
the following infinite derivation, obtained by infinite applications of rule (distr), proves that $\objectType{\C}{[\f_1{:}\bot,\f_2{:}\tr]}\leq\atr$ for all $\atr$:
$$
\dfrac{\dfrac{\vdots}{\objectType{\C}{[\f_1{:}\bot,\f_2{:}\tr]}\leq\atr}\qquad\dfrac{\vdots}{\objectType{\C}{[\f_1{:}\bot,\f_2{:}\tr]}\leq\atr}}
{\objectType{\C}{[\f_1{:}\bot,\f_2{:}\tr]}\leq\atr}
$$
Apparently this seems to be an unsound use of rule (distr) as it happens for rules ($\vee$R1) and ($\vee$R2) in the example above; however,
this is not the case, as we formally prove in the next section. Since $\objectType{\C}{[\f_1{:}\bot,\f_2{:}\tr]}\leq\atr$ and $\bot\leq\atr$ for all types $\atr$, then
$\bot\leq\objectType{\C}{[\f_1{:}\bot,\f_2{:}\intType]}$ and $\objectType{\C}{[\f_1{:}\bot,\f_2{:}\intType]}\leq\bot$ hold, that is, 
the two types are equivalent and, therefore, both represent the empty type. This result is not so surprising if we interpret the empty type
as the empty set of values, and we recall the similarity between records and Cartesian products, and the validity of the equation
$\emptyset\times V=\emptyset$. 

\begin{definition}
A derivation for $\tr_1\leq\tr_2$ is contractive iff it contains no sub-derivations built only with rules ($\vee$R1) and ($\vee$R2).
The subtyping relation $\tr_1\leq\tr_2$ holds iff there is a contractive derivation for it.
\end{definition}
In the following we use the term \emph{derivation} for contractive ones, unless explicitly specified.

\subsection{Interpretation of types}\label{sect-member}
We interpret types in a quite intuitive way, that is, as sets of values.
Values are all infinite terms coinductively defined by the following syntactic rules (where $i\in\Int$).
$$
\begin{array}{rcll}
\production{\val}{\num \vbar \objectType{\C}{[\f_1\mapsto\val_1,\ldots,\f_n\mapsto\val_n]} 
}
\end{array}
$$
As happens for object types, fields in object values are finite and
distinct, and their order is immaterial. Regular values correspond to finite, but cyclic, objects.

Membership of values to (the interpretation of) types is coinductively defined by the rules of Figure~\ref{fig-member}.
\begin{figure*}[t]
\begin{small}
\begin{center}
\begin{math}
\begin{array}{c}
\Rule{int}{}{i\in\intType}{}\qquad
\Rule{$\vee$L}{\val\in\tr_1}{\val\in\unionType{\tr_1}{\tr_2}}{}\qquad 
\Rule{$\vee$R}{\val\in\tr_2}{\val\in\unionType{\tr_1}{\tr_2}}{}
\\[4ex]
\Rule{obj}
{\val_1\in\tr_1,\ldots,\val_n\in\tr_n}
{\objectType{\C}{[\f_1\mapsto\val_1,\ldots,\f_n\mapsto\val_k,\ldots]}
\in\objectType{\C}{[\f_1{:}\tr_1,\ldots,\f_n{:}\tr_n]}}
{}
\end{array}
\end{math}
\caption{Rules defining membership}\label{fig-member}
\end{center}
\end{small}
\end{figure*}
All rules are intuitive. Note that an object value is allowed to belong
to an object type having less fields; this is expressed by the
ellipsis at the end of the values in the membership rule (obj). 

An analogous notion of contractive derivation has to be enforced also
for membership derivations. 
\begin{definition}
A derivation for $\val\in\tr$ is contractive iff it contains no
sub-derivations built only with membership rules ($\vee$R), and ($\vee$L).
The membership relation $\val\in\tr$ holds iff there is a contractive derivation for it.

The interpretation of type $\tr$ is denoted by $\intr{\tr}$ and defined by $\{\val \mid \val\in\tr \mbox{ holds} \}$.
\end{definition}

Before proving the main soundness theorem we show some examples of interpretations.

\paragraph{Example 1} If $\bot$ is the regular type s.t. $\bot=\unionType{\bot}{\bot}$, then $\intr{\bot}=\emptyset$. Indeed, the only
applicable rules are ($\vee$L) and ($\vee$R), hence only non contractive derivations can be built.

\paragraph{Example 2} If $\tr$ is the regular type s.t. $\tr=\unionType{\intType}{\tr}$, then $\intr{\tr}=\intr{\intType}=\Int$, that is,
$\tr$ and $\intType$ have the same interpretation. Indeed, all the
contractive derivations are obtained by applying $n$ times ($n\geq 0$)
rule ($\vee$R) (which is useless in this case),
then rule ($\vee$L) followed by (int):
$$
\dfrac
{}
{\dfrac{i\in\intType}{\dfrac{i\in\unionType{\intType}{\tr}}{\dfrac{\vdots}{i\in\unionType{\intType}{\tr}}}}}
$$
 
\paragraph{Example 3} Let us consider the infinite (but not regular) type $\tr_1$ defined by the following
infinite set of equations (where $\tr_1$ corresponds to $\var_0$):
\begin{small}
\begin{eqnarray*}
\var_0&=&\unionType{\avar_0}{\var_1}\\
&\ldots& \\
\var_n&=&\unionType{\avar_n}{\var_{n+1}}\\
&\ldots& \\
\avar_0&=&\objectType{\zero}{[\ ]}\\
\avar_1&=&\objectType{\scc}{[\prd{:}\avar_0]}\\
&\ldots& \\
\avar_{n+1}&=&\objectType{\scc}{[\prd{:}\avar_n]}\\
&\ldots&
\end{eqnarray*}
\end{small}
Let $\tr_2$ be the term s.t. $\tr_2=\unionType{\objectType{\zero}{[\ ]}}{\objectType{\scc}{[\prd{:}\tr_2]}}$.
Then $\intr{\tr_1}\subsetneq\intr{\tr_2}$; indeed, it is easy to show
that $\intr{\tr_1}$ is the set of all objects representing 
natural numbers, and that such values belong to $\intr{\tr_2}$ as well (all derivations are finite, hence trivially contractive),
whereas the value $\val_\infty$ s.t. $\val_\infty=\objectType{\scc}{[\prd\mapsto \val_\infty]}$ belongs to $\tr_2$, but not to $\tr_1$.  
Indeed, the following contractive and regular derivation can be built by
alternatively applying rules ($\vee$R) and (obj) infinite times.
$$
\dfrac
{\dfrac
  {\dfrac
  {\vdots}
  {\val_\infty\in\tr_2}
  }
  {\val_\infty\in\objectType{\scc}{[\prd{:}\tr_2]}
 }
}
{\val_\infty\in\tr_2}
$$
Finally, it is not difficult to prove that the only derivation for
$\val_\infty\in\tr_1$ is not contractive, since it can be obtained by infinitely applying 
rule ($\vee$R); therefore  $\val_\infty\not\in\tr_1$.

\section{Soundness}\label{sect-sound}
We now prove that the definition of $\leq$ is sound  w.r.t. 
containment between type interpretations.
The proof of soundness is based on the following lemma.
\begin{lemma}\label{lemma-one}
If $\tr$ is an object type s.t. $\tr\leq\atr$ and $\val\in\tr$,
then there exists an object type $\tr'$ (not necessarily equal to $\tr$) s.t. 
$\val\in\tr'$, and s.t. there exists a derivation for $\tr'\leq\atr$ whose first applied rule is ($\vee$R1), ($\vee$R2) or (obj). 
\end{lemma}
\proof
The proposed proof is constructive, since it shows that the derivation for $\tr'\leq\atr$ is just a sub-derivation of
the derivation for $\tr\leq\atr$, and that
the derivation for $\val\in\tr'$ can be easily built from the derivation for $\val\in\tr$.

Let $\tr=\objectType{\C}{[\f_1{:}\tr_1,\ldots,\f_n{:}\tr_n]}$, by membership rule (obj)
$\val=\objectType{\C}{[\f_1\mapsto\val_1,\ldots,\f_n\mapsto\val_n,\ldots]}$; furthermore, the corresponding derivation has the following shape:
$$
\dfrac
    {\begin{array}{ccc}
        \dfrac
            {\dfrac
              {\dfrac
                {\vdots}
                {\val_1\in\tr'_1}
              }
              {\begin{array}{ccc}& . & \\[-1.5ex] & . & \scriptstyle k_1 \\[-1.5ex]& . &  \end{array}}
            }
            {\val_1\in\tr_1} &
        \dots &    
        \dfrac
            {\dfrac
              {\dfrac
                {\vdots}
                {\val_n\in\tr'_n}
              }
              {\begin{array}{ccc}& . & \\[-1.5ex] & . & \scriptstyle k_n \\[-1.5ex]& . &  \end{array}}
            }
            {\val_n\in\tr_n}            
    \end{array}}
    {\val\in\objectType{\C}{[\f_1{:}\tr_1,\ldots,\f_n{:}\tr_n]}}
$$
where $\tr'_1,\ldots,\tr'_n$ are not union types, and are obtained after repeatedly applying
rules ($\vee$L) or ($\vee$R) $k_1,\ldots,k_n$ times respectively. We know that all $k_i$
are finite, otherwise the derivation would not be contractive. The proof proceeds by induction
on  $m=\sum_{i\in 1\ldots n}k_i$.   

If $m=0$, then all $\tr_1,\ldots,\tr_n$ are not union types. If
$\atr=\intType$, then there are no applicable subtyping rules
and the claim trivially holds since the hypothesis is not satisfied; if $\atr$ is either a union or an object type,  
then the only applicable subtyping rules
are ($\vee$R1), ($\vee$R2) or (obj), therefore we easily conclude with $\tr'=\tr$. 
If $m>0$ and the derivation is obtained by applying rule\footnote{If one between ($\vee$R1), ($\vee$R2), and (obj) has been applied, then the conclusion is straightforward
as for $m=0$.} (distr), then $\tr_1=\unionType{\tr_a}{\tr_b}$, that
is, $\tr=\objectType{\C}{[\f_1{:}\unionType{\tr_a}{\tr_b},\ldots,\f_n{:}\tr_n]}$.
Furthermore, in the derivation for $\val\in\tr$, the first applied rule of the sub-derivation for
$\val_1\in\unionType{\tr_a}{\tr_b}$ is either ($\vee$L) or ($\vee$R). If ($\vee$L) has been applied 
(the other case is completely symmetric), then a derivation for $\val\in\objectType{\C}{[\f_1{:}\tr_a,\ldots,\f_n{:}\tr_n]}$ can be obtained from that
of $\val\in\tr$, by simply removing the application of rule ($\vee$L)
for $\val_1\in\unionType{\tr_a}{\tr_b}$, as depicted in Figure~\ref{fig-lemma}.
Therefore in such derivation $\sum_{i\in 1\ldots n}k_i=m-1$. Finally, since rule (distr) has been applied, we know that 
$\objectType{\C}{[\f_1{:}\tr_a,\ldots,\f_n{:}\tr_n]}\leq \atr$, hence we can conclude by inductive hypothesis. 

As a final remark, note that the construction of $\tr'$ and of the
derivations for $\tr'\leq\atr$ and $\val\in\tr'$ are uniquely determined by the derivations 
for $\tr\leq\atr$ and $\val\in\tr$. Therefore, the proof
of the lemma shows that there exists a function $\lemmafun$ s.t. if
$\dv_1$ and $\dv_2$ are derivations 
for $\tr\leq\atr$ and $\val\in\tr$, respectively, with $\tr$ object
type, then $\lemmafun(\dv_1,\dv_2)$ returns $(\dv_3,\dv_4)$ s.t. 
$\dv_3$ and $\dv_4$ are derivations for $\tr'\leq\atr$ and
$\val\in\tr'$, respectively, where $\tr'$ is an object type,  $\dv_3$ is a sub-derivation of
$\dv_1$ where the first applied
rule is ($\vee$R1), ($\vee$R2) or (obj), and $\dv_4$ is obtained by $\dv_2$ by replacing some node and removing some applications of rules ($\vee$L) and ($\vee$R). 
\qed
\begin{figure*}[t]
\begin{small}
\begin{center}
$$
\begin{array}{ccc}
\dfrac
    {\begin{array}{ccc}
        {\mbox{\tiny ($\vee$L)}}\dfrac
            {\dfrac{
                \dfrac
                    {\dfrac
                      {\vdots}
                      {\val_1\in\tr'_1}
                    }
                    {\begin{array}{ccc}& . & \\[-1.5ex] & . & \scriptstyle k_1-1 \\[-1.5ex]& . &  \end{array}}
              }
              {\val_1\in\tr_a}
            }
            {\val_1\in\unionType{\tr_a}{\tr_b}} &
            \dots &    
            \dfrac
                {\dfrac
                  {\dfrac
                    {\vdots}
                    {\val_n\in\tr'_n}
                  }
                  {\begin{array}{ccc}& . & \\[-1.5ex] & . & \scriptstyle k_n \\[-1.5ex]& . &  \end{array}}
                }
                {\val_n\in\tr_n}            
    \end{array}}
    {\val\in\objectType{\C}{[\f_1{:}\unionType{\tr_a}{\tr_b},\ldots,\f_n{:}\tr_n]}} &
\Longrightarrow &
\dfrac
    {\begin{array}{ccc}
        \dfrac
            {\dfrac
              {\dfrac
                {\vdots}
                {\val_1\in\tr'_1}
              }
              {\begin{array}{ccc}& . & \\[-1.5ex] & . & \scriptstyle k_1-1 \\[-1.5ex]& . &  \end{array}}
            }
            {\val_1\in\tr_a} &
        \dots &    
        \dfrac
            {\dfrac
              {\dfrac
                {\vdots}
                {\val_n\in\tr'_n}
              }
              {\begin{array}{ccc}& . & \\[-1.5ex] & . & \scriptstyle k_n \\[-1.5ex]& . &  \end{array}}
            }
            {\val_n\in\tr_n}            
    \end{array}}
    {\val\in\objectType{\C}{[\f_1{:}\tr_a,\ldots,\f_n{:}\tr_n]}} 
\end{array}
$$
\caption{Transformation of derivations in proof of lemma~\ref{lemma-one}}\label{fig-lemma}
\end{center}
\end{small}
\end{figure*}

\begin{theorem}[Soundness]\label{theo-sound}
For all $\tr_1,\tr_2$, if $\tr_1\leq\tr_2$, then $\intr{\tr_1}\subseteq\intr{\tr_2}$.
\end{theorem}
\proof
The claim can be put in the following equivalent form: 
for all $\tr_1,\tr_2$,$\val$ if $\tr_1\leq\tr_2$, $\val\in\tr_1$ then $\val\in\tr_2$.

The proof is constructive, since it coinductively defines a function $\fun$ from derivations for
$\tr_1\leq\tr_2$ and $\val\in\tr_1$ to derivations for $\val\in\tr_2$.
The definition of $\fun$ is given by cases on the first applied subtyping rule of the derivation for 
$\tr_1\leq\tr_2$.

\paragraph{Rule (int)} 
$\fun\left(\mbox{\scriptsize (int)}\dfrac{}{\intType\leq\intType},\mbox{\scriptsize (int)}\dfrac{}{i\in\intType}\right)={\mbox{\scriptsize (int)}}\frac{}{i\in\intType}$.

\paragraph{Rule ($\vee$R1)} 
$\fun\left(\mbox{\scriptsize
  ($\vee$R1)}\dfrac{\dv_1}{\tr_1\leq{\unionType{\atr_1}{\atr_2}}},\dv_2\right)=
\mbox{\scriptsize ($\vee$L)}\dfrac{\fun(\dv_1,\dv_2)}{\val\in\unionType{\atr_1}{\atr_2}}$,
where $\dv_1$ is a derivation for $\tr_1\leq\atr_1$, and
$\dv_2$ is a derivation for $\val\in\tr_1$.

\paragraph{Rule ($\vee$R2)}
$\fun\left(\mbox{\scriptsize
  ($\vee$R2)}\dfrac{\dv_1}{\tr_1\leq{\unionType{\atr_1}{\atr_2}}},\dv_2\right)=
\mbox{\scriptsize ($\vee$R)}\dfrac{\fun(\dv_1,\dv_2)}{\val\in\unionType{\atr_1}{\atr_2}}$,
where $\dv_1$ is a derivation for $\tr_1\leq\atr_2$, and
$\dv_2$ is a derivation for $\val\in\tr_1$.

\paragraph{Rule ($\vee$L)} There are two sub-cases, depending on the
shape of the derivation for $\val\in\tr_2$:\\[2ex]
$\fun\left(\mbox{
  \scriptsize($\vee$L)}\dfrac{\dv_1\quad\dv_2}{\unionType{\atr_1}{\atr_2}\leq\tr_2},
\mbox{
  \scriptsize($\vee$L)}\dfrac{\dv_3}{\val\in\unionType{\atr_1}{\atr_2}}\right)=\fun(\dv_1,\dv_3)$\\[2ex]
$\fun\left(\mbox{
  \scriptsize($\vee$L)}\dfrac{\dv_1\quad\dv_2}{\unionType{\atr_1}{\atr_2}\leq\tr_2},
\mbox{
  \scriptsize($\vee$R)}\dfrac{\dv_4}{\val\in\unionType{\atr_1}{\atr_2}}\right)=\fun(\dv_2,\dv_4)$\\[2ex]
 In this case $\dv_1$ and $\dv_2$ are derivations for
 $\atr_1\leq\tr_2$ and $\atr_2\leq\tr_2$, respectively, whereas
$\dv_3$ and $\dv_4$
are derivations for $\val\in\atr_1$ and $\val\in\atr_2$, respectively.

\paragraph{Rule (obj)}\hspace*{\fill}\\[2ex]
$
\begin{array}{l}
\fun \left(  
\begin{array}{l}
  \mbox{\scriptsize
  (obj)}\dfrac{\dv_1,\ldots,\dv_n}{\objectType{\C}{[\f_1{:}\atr_1,\ldots,\f_n{:}\atr_n,\ldots]}\leq\objectType{\C}{[\f_1{:}\atr'_1,\ldots,\f_n{:}\atr'_n]}}, \\[3ex]
 \mbox{
  \scriptsize(obj)}\dfrac{\dv'_1,\ldots,\dv'_n,\ldots} 
 {\objectType{\C}{[\f_1\mapsto\val_1,\ldots,\f_n\mapsto\val_n,\ldots]}\in\objectType{\C}{[\f_1{:}\atr_1,\ldots,\f_n{:}\atr_n,\ldots]}}
 \end{array}
\right)= \\[8ex]
\qquad\qquad\mbox{
  \scriptsize(obj)}\dfrac{\fun(\dv_1,\dv'_1),\ldots,\fun(\dv_n,\dv'_n)}{\objectType{\C}{[\f_1\mapsto\val_1,\ldots,\f_n\mapsto\val_n,\ldots]}\in\objectType{\C}{[\f'_1{:}\atr_1,\ldots,\f'_n{:}\atr_n]}}
\end{array}
$\\[2ex] 
where $\dv_1,\ldots,\dv_n$ are derivations for $\atr_1\leq\atr'_1,\ldots,\atr_n\leq\atr'_n$, respectively,  whereas
$\dv'_1,\ldots,\dv'_n$ are derivations for
$\val_1\in\atr_1,\ldots,\val_n\in\atr_n$, respectively. 

The derivation for
$\objectType{\C}{[\f_1\mapsto\val_1,\ldots,\f_n\mapsto\val_n,\ldots]}\in\objectType{\C}{[\f_1{:}\atr_1,\ldots,\f_n{:}\atr_n,\ldots]}$
contains ellipses in the right hand side of the sub-derivations $\dv'_1,\ldots,\dv'_n$
and of the fields of both the value and the type. Their meaning is
that there may be other entities in the derivation which, however, can be
omitted, since the definition of $\fun$ does not depend on them.  

\paragraph{Rule (distr)} In this case the hypotheses of
lemma~\ref{lemma-one} are verified, therefore we can use the
function $\lemmafun$ defined in the proof of the lemma:
$$\fun(\dv_1,\dv_2)=\fun(\lemmafun(\dv_1,\dv_2))$$ where $\dv_1$ is a
derivation for $\tr_1\leq\tr_2$ whose first applied rule is (distr),
hence $\tr_1$ is an object type,
and $\dv_2$ is a derivation for $\val\in\tr_1$. 
According to the proof of the lemma, $\lemmafun(\dv_1,\dv_2)$ returns
$(\dv_3,\dv_4)$ s.t.  
$\dv_3$ and $\dv_4$ are derivations for $\tr\leq\tr_2$ and
$\val\in\tr$,  $\tr$ is an object type, and the first applied
rule of $\dv_3$ is ($\vee$R1), ($\vee$R2), or
(obj). Therefore case (distr) is delegated to one of the three cases
($\vee$R1), ($\vee$R2), (obj) specified above.

Now the remaining part of the proof is showing that $\fun$ is
well-defined. Since $\fun$ is defined coinductively, we need to prove
that $\fun$ is a function, that is, it cannot return two different 
derivations when applied to the same arguments.  
To show this, we first prove the following property.
\paragraph{Property (*)} If $\dv_1$ and $\dv_2$ are derivations for
$\tr_1\leq\tr_2$ and $\val\in\tr_1$, respectively, and $(\dv_1,\dv_2)$ matches cases ($\vee$L) or
(distr) of the definition of $\fun$, then there always exist $\dv_3$ and $\dv_4$
s.t. for any derivation $\dv$ returned by $\fun(\dv_1,\dv_2)$, the
following facts hold: $\dv=\fun(\dv_3,\dv_4)$, there exists $\tr$
s.t. $\dv_3$ and $\dv_4$ are derivations for $\tr\leq\tr_2$ and
$\val\in\tr$, respectively, and  $(\dv_3,\dv_4)$
matches one between (int), ($\vee$R1), ($\vee$R2), and (obj) cases. 
\paragraph{Proof of (*):} 
It is immediate to prove that if $\dv_1$ and
$\dv_2$ are derivations for $\tr_1\leq\tr_2$ and $\val\in\tr_1$,
respectively, then there always exists one and only one case matching
$(\dv_1,\dv_2)$ in the definition of $\fun$.  
If $(\dv_1,\dv_2)$ matches case (distr), then by
lemma~\ref{lemma-one} we know that $\lemmafun$ is defined on
$(\dv_1,\dv_2)$, and returns $(\dv_3,\dv_4)$ s.t.  
$\dv_3$ and $\dv_4$ are derivations for $\tr\leq\tr_2$ and
$\val\in\tr$, where $\tr$ is an object type, and the first applied
rule of $\dv_3$ is ($\vee$R1), ($\vee$R2) or
(obj). Now, since $(\dv_1,\dv_2)$ cannot match any other case,
by definition of $\fun$ we can conclude that  
for any $\dv$ returned by $\fun(\dv_1,\dv_2)$, the equality
$\dv=\fun(\lemmafun(\dv_1,\dv_2))=\fun(\dv_3,\dv_4)$ must hold.

If $(\dv_1,\dv_2)$ matches case
($\vee$L), then we proceed by induction on the number $n$ of
contiguous applications of membership rules ($\vee$L) and ($\vee$R)
with which derivation $\dv_2$ starts. We know that such $n$ is finite,
otherwise $\dv_2$ would not be contractive. The basis if for $n=1$,
since for $n=0$  the pair $(\dv_1,\dv_2)$ would not match case
($\vee$L); for simplicity, let us assume that $\dv_2$
starts with the application of rule ($\vee$L), that is, the first
sub-case applies (the other sub-case is
symmetric). Then we know that $\dv_1$ and $\dv_2$ have the following shape:
$$
\dv_1=\mbox{ \scriptsize($\vee$L)}\dfrac{\dv_3\quad\dv'_3}{\unionType{\tr}{\tr'}\leq\tr_2}\qquad
\dv_2=\mbox{ \scriptsize($\vee$L)}\dfrac{\dv_4}{\val\in\unionType{\tr}{\tr'}}
$$
where $\dv_3$ and $\dv_4$ are derivations for $\tr\leq\tr_2$ and $\val\in\tr$, respectively.
Since $(\dv_1,\dv_2)$ cannot match any other case,
by definition of $\fun$ we have that 
for any $\dv$ returned by $\fun(\dv_1,\dv_2)$, the equality
$\dv=\fun(\dv_3,\dv_4)$ must hold.
Finally, $(\dv_3,\dv_4)$ must match
some case of the definition of $\fun$, but such case cannot be
($\vee$L); indeed, $n=1$ and, therefore, $\tr$ cannot be a union type.
In case $(\dv_3,\dv_4)$ matches case (distr), we can
apply\footnote{This is possible because proof of case (distr) does not
depend on proof of case ($\vee$L).} the result 
already proved for that case.
The inductive step is a direct consequence of the inductive hypothesis
and of the fact that if $\dv_2$ starts with $n+1$ consecutive
applications of rules ($\vee$L) and ($\vee$R), then $\dv_4$ starts with $n$ consecutive
applications of rules ($\vee$L) and ($\vee$R).  
 
We can now prove the following property.  
\paragraph{$\fun$ is deterministic:} For all $\dv_1, \dv_2, \dv,
\dv'$, if $\fun(\dv_1,\dv_2)=\dv$ and $\fun(\dv_1,\dv_2)=\dv'$, then
$\dv=\dv'$.

We prove that $\dv=\dv'$ by induction on the height of the finite trees approximating $\dv$ and $\dv'$, that is, we
show that all paths of $\dv$ starting from its root are equal to the
paths of $\dv'$ starting from its root, for all the
lengths\footnote{Recall that the path from the root to a given node is
always finite, even when the tree is infinite.} of the
paths. The basis consists in proving that $\dv$ and $\dv'$ have the
same root and start with the same rule application (that is, the path
length is 0). This comes directly from the definition of $\fun$ for the cases
(int), ($\vee$R1), ($\vee$R2), and (obj), from the fact that all cases
are disjoint, and from property (*) (which deals with the two remaining cases).
The inductive step is derived from these same facts, from the
inductive hypothesis, and from the standard definition of path length. 

\paragraph{$\fun$ returns contractive derivations:} 
If $\dv_1$ and $\dv_2$ are derivations for $\tr_1\leq\tr_2$, $\val\in\tr_1$, respectively, then
$\fun(\dv_1,\dv_2)$ is defined and is a derivation for
$\val\in\tr_2$. 

First, we recall that the definition of $\fun$ covers all possible
cases, then $\fun$ is always defined on $(\dv_1,\dv_2)$.
Then we show that the tree returned by $\fun$ is always a derivation,
and finally we prove that all returned derivations are contractive.
To prove that all returned trees are derivations, we first observe that $\fun$ always returns a tree having shape
$\frac{\dv}{\val\in\tr_2}$. Again, this comes directly from the definition of $\fun$ for the cases
(int), ($\vee$R1), ($\vee$R2), and from property (*) (which deals with
the two remaining cases).
Then the proof proceeds by induction 
on the height of the finite derivations approximating
$\fun(\dv_1,\dv_2)$. That is, we prove that every node whose
distance\footnote{Where the distance is the length of the path from
  the node to the root.}
from the root has length less or equal than $n$ is obtained with  
a correct rule instantiation, for all $n$.
The basis (for $n=0$) comes directly from the definition of $\fun$ for the cases
(int), ($\vee$R1), ($\vee$R2), and from property (*). Let us see case ($\vee$R1)
as an example. In this case we know that $\fun(\dv_1,\dv_2)=
\mbox{\scriptsize ($\vee$L)}\frac{\fun(\dv_3,\dv_4)}{\val\in\unionType{\atr_1}{\atr_2}}$,
where $\dv_3$ is a derivation for $\tr_1\leq\atr_1$, and
$\dv_3$ is a derivation for $\val\in\tr_1$, therefore the root of 
$\fun(\dv_3,\dv_4)$ is $\val\in\atr_1$, hence $\unionType{\atr_1}{\atr_2}$ is obtained with
a correct instantiation of rule ($\vee$L). 
The inductive step is derived from the definition of $\fun$ for the cases
(int), ($\vee$R1), ($\vee$R2), from property (*), from the
inductive hypothesis, and from the standard definition of path length. 
 
We conclude the proof by showing that if $\dv_1$ and $\dv_2$ are contractive, then $\fun(\dv_1,\dv_2)$ is contractive as well.
By contradiction, let us assume that the returned derivation is not contractive, that is,
there exists a sub-derivation containing just applications of memberships rules ($\vee$L) and ($\vee$R).
Since ($\vee$R1) and ($\vee$R2) are the only two cases
where an application of membership rule ($\vee$L) or ($\vee$R) is
added to the returned derivation, and  cases ($\vee$L) and (distr) may be 
defined in terms of cases ($\vee$R1) and ($\vee$R2), then such a sub-derivation
can be built by applying only cases ($\vee$R1), ($\vee$R2), ($\vee$L) and (distr) of the definition of $\fun$.
Now we observe that if case (distr) occurs, then, by definition of $\lemmafun$ given in lemma~\ref{lemma-one}, and by definition
of cases ($\vee$R1) and ($\vee$R2), only cases ($\vee$R1) and ($\vee$R2) may occur afterwards; but this means that
$\dv_1$ contains a sub-derivation built only with rules ($\vee$R1) and ($\vee$R2), that is, $\dv_1$ is not contractive, which is in contradiction 
with the hypothesis.
If case (distr) does not occur, and case ($\vee$L) occurs infinite times, then
by definition of cases ($\vee$R1), ($\vee$R2), and ($\vee$L), we deduce that
$\dv_2$ is not contractive, against the hypothesis.
The last possibility is when case (distr) does not occur, and case ($\vee$L) occurs only a finite numbers of time; but
this necessarily means that at a certain point only cases ($\vee$R1) and ($\vee$R2) may occur, 
that is, $\dv_1$ is not contractive, which is in contradiction 
with the hypothesis.
\qed

\section{A complete characterization of the empty type}\label{sect-complete}
We have already shown in Section~\ref{sect-subtype} that $\objectType{\C}{[\f_1{:}\bot,\f_2{:}\tr]}\leq\bot$, where $\bot$ is the empty type, that is,
the type s.t. $\bot=\unionType{\bot}{\bot}$; therefore, $\bot$ and $\objectType{\C}{[\f_1{:}\bot,\f_2{:}\tr]}$ are equivalent.
In fact, besides $\objectType{\C}{[\f_1{:}\bot,\ldots]}$, there are infinitely many other types equivalent to $\bot$, namely, all object types
``containing'' $\bot$.

For instance, the type $\tr=\objectType{\C_1}{[\f{:}\objectType{\C_2}{[\g{:}\bot]}]}$ is s.t.
$\intr{\tr}=\emptyset$. Unfortunately, $\tr\leq\bot$ is not derivable from the rules in Figure~\ref{fig-subtype}.
Indeed, all possible derivations can be built by only applying rules ($\vee$R1) and ($\vee$R2), and are, therefore, not contractive.
To overcome this problem, we introduce a rule explicitly dealing with all types equivalent to the empty type.
In order to do that, we would need a predicate 
$\emptyType{\tr}$ defining all types $\tr$ equivalent to $\bot$. However, the complementary
predicate $\notEmptyType{\tr}$ turns out to be more convenient, because of its strong similarity with the membership relation;
indeed, a type $\tr$ is not equivalent to the empty type iff there exists a value $\val$ s.t. $\val\in\tr$ holds. In this way, it is quite straightforward
to prove that the predicate  $\notEmptyType{\tr}$ is sound and complete w.r.t. our type interpretation.
Hence, our new subtyping rule is defined as follows.
$$
\Rule
{empty}
{}
{\tr_1\leq\tr_2}
{\mbox{$\negNotEmptyType{\tr_1}$}}
$$   
The definition of $\notEmptyType{\tr}$ is quite straightforward.
$$
\begin{small}
\begin{array}{c}
\Rule
{$\uparrow\vee$L}
{\notEmptyType{\tr_1}}
{\notEmptyType{\unionType{\tr_1}{\tr_2}}}
{}
\quad\quad
\Rule
{$\uparrow\vee$R}
{\notEmptyType{\tr_2}}
{\notEmptyType{\unionType{\tr_1}{\tr_2}}}
{}
\quad\quad
\Rule
{$\uparrow$ int}
{}
{\notEmptyType{\intType}}
{}
\quad\quad
\Rule{$\uparrow$ obj}
{\notEmptyType{\tr_1},\ldots,\notEmptyType{\tr_n}}
{\notEmptyType{\objectType{\C}{[\f_1{:}\tr_1,\ldots,\f_n{:}\tr_n]}}}
{}
\end{array}
\end{small}
$$   
As usual, all derivations have to be contractive, hence they cannot contain sub-derivations obtained by only applying rules
($\uparrow\vee$L) and ($\uparrow\vee$R). 

Note that if we restrict ourselves to regular types, then the definition of 
$\notEmptyType{}$ can be turned into the following algorithm specified in pseudo-Java code.
\begin{center}
\begin{lstlisting}[frame=single]
boolean not_empty(type $\tr$,stack path) {
  if($\tr$.is_visited()) 
    return path.is_contractive($\tr$);
  else {
    $\tr$.set_visited();
    switch($\tr$) {
        case $\intType$: return true;
        case $\unionType{\tr_1}{\tr_2}$:
          path.push($\tr$);
          if(not_empty($\tr_1$,path)) {
            path.pop(); 
            return true;
          }
          res=not_empty($\tr_2$,path);
          path.pop();
          return res;
        case $\objectType{\C}{[\f_1{:}\tr_1,\ldots,\f_n{:}\tr_n]}$:
          path.push($\tr$);
          for $i\in 1,\ldots,n$ {
            if (!not_empty($\tr_i$,path)) {
              path.pop();
              return false;
            }
          } 
          path.pop();
          return true;
      }
  }
}
\end{lstlisting}
\end{center} 
The argument $\tr$ is the type to be inspected, whereas \lstinline{path} contains the stack of visited nodes, which 
must be initially empty. Such a stack is used for checking that the found derivation is contractive.
Methods \lstinline{is_visited} and \lstinline{set_visited} are used to keep track of visited terms, which correspond to nodes in a graph. 
If we end up with an already visited type, then we have an infinite regular path that, however, has to be 
contractive, otherwise the corresponding derivation is not valid: method \lstinline{is_contractive} checks whether there is 
an object type in the sub-path of \lstinline{path} from $\tr$ to the top of the stack. 
The time complexity of the algorithm is linear in the number of edges of the graph representing the term, providing that
\lstinline{is_contractive} has a constant time\footnote{This can be achieved by associating a position with each node in the path, and by recording the minimum position
$p$ s.t. all paths starting from a node whose position is greater than $p$ are non contractive.} complexity.
  
We can now prove that the definition of $\notEmptyType{}$ is sound and complete w.r.t. the interpretation of types.
\begin{theorem}[Soundness of $\notEmptyType{\tr}$]
If $\notEmptyType{\tr}$, then $\intr{\tr}\neq\emptyset$.
\end{theorem}
\proof
Similarly to the proof of Theorem~\ref{theo-sound}, we coinductively define a function $\fun$ mapping derivations
for $\notEmptyType{\tr}$ to derivations for $\val\in\tr$, for a fixed value $\val$:
$$
\begin{array}{l}
\fun\left(\mbox{\scriptsize (int)}\dfrac{} {\notEmptyType{\intType}}\right)=\mbox{\scriptsize (int)}\dfrac{}{0\in\intType} \qquad
\fun\left(\mbox{\scriptsize ($\vee$L)}\dfrac{\dv}{\notEmptyType{\unionType{\tr_1}{\tr_2}}}\right)=\mbox{\scriptsize ($\vee$L)}\dfrac{\fun(\dv)}{\val\in\unionType{\tr_1}{\tr_2}} \\[2ex]
\fun\left(\mbox{\scriptsize ($\vee$R)}\dfrac{\dv}{\notEmptyType{\unionType{\tr_1}{\tr_2}}}\right)=\mbox{\scriptsize ($\vee$R)}\dfrac{\fun(\dv)}{\val\in\unionType{\tr_1}{\tr_2}} \\[2ex]
\fun\left(\mbox{\scriptsize (obj)}\dfrac{\dv_1,\ldots,\dv_n}{\notEmptyType{\objectType{\C}{[\f_1{:}\tr_1,\ldots,\f_n{:}\tr_n]}}}\right)=
\mbox{\scriptsize (obj)}\dfrac{\fun(\dv_1),\ldots,\fun(\dv_n)}{\objectType{\C}{[\f_1\mapsto\val_1,\ldots,\f_n\mapsto\val_n]}\in\objectType{\C}{[\f_1{:}\tr_1,\ldots,\f_n{:}\tr_n]}} 
\end{array}
$$
Not that $\fun$ fully preserves the shape of derivations, in the sense that only the derived judgments change.
Using a similar, but simpler, proof scheme as adopted for Theorem~\ref{theo-sound}, it is possible to prove that 
the above definition corresponds to a function $\fun$ s.t. for all derivations $\dv$ for $\notEmptyType{\tr}$, $\fun(\dv)$ is a derivation for $\val\in\tr$, for a certain $\val$.
\qed

\begin{theorem}[Completeness of $\notEmptyType{\tr}$]\label{theo-complete-not-empty}
If $\intr{\tr}\neq\emptyset$, then $\notEmptyType{\tr}$.
\end{theorem}
\proof
The proof is similar to that for soundness, except that here the function definition is even simpler, since it basically 
forgets the value $\val$ in the membership judgment.
$$
\begin{array}{l}
\fun\left(\mbox{\scriptsize (int)}\dfrac{}{\val\in\intType}\right)=\mbox{\scriptsize (int)}\dfrac{}{\notEmptyType{\intType}} \qquad
\fun\left(\mbox{\scriptsize ($\vee$L)}\dfrac{\dv}{\val\in\unionType{\tr_1}{\tr_2}}\right)=\mbox{\scriptsize ($\vee$L)}\dfrac{\fun(\dv)}{\notEmptyType{\unionType{\tr_1}{\tr_2}}} \\[2ex]
\fun\left(\mbox{\scriptsize ($\vee$R)}\dfrac{\dv}{\val\in\unionType{\tr_1}{\tr_2}}\right)=\mbox{\scriptsize ($\vee$R)}\dfrac{\fun(\dv)}{\notEmptyType{\unionType{\tr_1}{\tr_2}}} \\[2ex]
\fun\left(\mbox{\scriptsize (obj)}\dfrac{\dv_1,\ldots,\dv_n}{\objectType{\C}{[\f_1\mapsto\val_1,\ldots,\f_n\mapsto\val_n]}\in\objectType{\C}{[\f_1{:}\tr_1,\ldots,\f_n{:}\tr_n]}}\right)=
\mbox{\scriptsize (obj)}\dfrac{\fun(\dv_1),\ldots,\fun(\dv_n)}{\notEmptyType{\objectType{\C}{[\f_1{:}\tr_1,\ldots,\f_n{:}\tr_n]}}} 
\end{array}
$$
\qed

This final result allows us to fully reuse the proof of Theorem~\ref{theo-sound} to show that subtyping remains sound w.r.t. containment between type interpretations, if rule (empty)
is added.
\begin{corollary}
The subtyping relation coinductively defined by rules in Figure~\ref{fig-subtype}, and by rule (empty) is sound w.r.t.  containment between type interpretations.
\end{corollary}
\proof
It suffices considering the same function $\fun$ defined in proof of Theorem~\ref{theo-sound}, since the new case (empty) cannot occur; indeed, there exist no derivations 
$\dv_1$ and $\dv_2$ for $\tr_1\leq\tr_2$ and $\val\in\tr_2$, respectively, s.t. the first applied rule of $\dv_1$ is (empty), because, by the side condition of rule (empty), 
$\negNotEmptyType{\tr_1}$, and, hence, by Theorem~\ref{theo-complete-not-empty}, $\intr{\tr_1}=\emptyset$.  
\qed

\section{Conclusion}\label{sect-conclu}

We have studied a subtyping relation on coinductive terms built on object and union types constructors, by providing
a quite natural interpretation based on a membership relation of values to types, and proved that such a relation is sound
w.r.t. containment between type interpretations.

This study has allowed us to improve the original definition of subtyping \cite{AnconaLagorio09} in two different directions:
\begin{itemize}
\item Contractiveness was too restrictive, since no derivations built only with ($\vee$R1), ($\vee$R2), and (distr) rules were allowed, whereas
the type interpretation and the corresponding proof of soundness given here have shown that no restrictions on rule (distr) is ever needed.
Consequently, the subtyping relation can be implemented more directly, since, rules ($\vee$R1) and ($\vee$R2) have only one premise, in contrast with (distr), and, therefore,
checking contractiveness of derivations is simpler.

\item The definition did not consider all possible representations of the empty type.
Consequently a corresponding new rule has been added, and a sound and complete characterization of all representations 
of the empty type has been provided; when restricted to regular types, such a characterization directly provides an algorithm
for checking whether the interpretation of a type is empty. The time complexity of the algorithm is linear 
in the number of edges of the graph representing the term.
\end{itemize}

\bibliographystyle{eptcs}
\bibliography{ALLMY}

\appendix
\section{Appendix: Horn clauses generated by the code examples in Section~\ref{sect-ex}}

The last clauses of \lstinline{has_field} and \lstinline{has_meth} are essential for correctly dealing
with inherited fields and methods, respectively, even though they could be safely omitted here, since
classes \lstinline{Zero} and \lstinline{Succ} do not inherit any field or method. Note that we have used negation just 
for brevity, but it can always be omitted by defining the trivial predicates \lstinline{not_dec_field} and
\lstinline{not_dec_meth}, since \lstinline{dec_field} and \lstinline{dec_meth} are simply defined by a collection of ground facts.  

Finally, note that the definition of predicate \lstinline{field_acc} (for field access) depends on the predicate
\lstinline{rec_acc} (for record access) which is defined by a single clause containing just a singleton record;
this is correct thanks to subsumption and subtyping on record types. For instance, since the goal 
\lstinline[language=Prolog]{rec_acc([f1:int],f1,int)} is derivable, and \lstinline[language=Prolog]{[f1:int,f2:obj(c,[])]} is a subtype of 
\lstinline[language=Prolog]{[f1:int]}, then \lstinline[language=Prolog]{rec_acc([f1:int,f2:obj(c,[])],f1,int)} is derivable as well, by subsumption.    

\begin{lstlisting}[language=Prolog]
class(object).
class(zero).
class(succ).
extends(zero,object).
extends(succ,object).
subclass(X,X) $\leftarrow$ class(X).
subclass(X,object) $\leftarrow$ class(X).
subclass(X,Y) $\leftarrow$ extends(X,Z),subclass(Z,Y).
field_acc(obj(C,R),F,T) $\leftarrow$ has_field(C,F),rec_acc(R,F,T).
field_acc(T1$\vee$T2,F,FT1$\vee$FT2) $\leftarrow$  field_acc(T1,F,FT1),field_acc(T1,F,FT1).
rec_acc([F:T],F,T).
invoke(obj(C,R),M,A,RT) $\leftarrow$  has_meth(C,M,[obj(C,R)|A],RT).
invoke(T1$\vee$T2,M,A,RT1$\vee$RT2) $\leftarrow$  invoke(T1,M,A,RT1),invoke(T2,M,A,RT2).
new(object,[],obj(object,[])).
new(zero,[],obj(zero,R)) $\leftarrow$ extends(zero,P),new(P,[],obj(P,R)).
new(succ,[N],obj(succ,[pred:N|R])) $\leftarrow$ extends(succ,P),new(P,[],obj(P,R)).
dec_field(succ,pred).
has_field(C,F) $\leftarrow$ dec_field(C,F).
has_field(C,F) $\leftarrow$ extends(C,P),has_field(P,F),$\neg$dec_field(C,F).
dec_meth(zero,add).
dec_meth(succ,add).
has_meth(zero,add,[This,N],N).
has_meth(succ,add,[This,N],R) $\leftarrow$ field_acc(This,pred,P),new(succ,[N],S),
                                           invoke(P,add,[S],R).
has_meth(C,M,A,R) $\leftarrow$ extends(C,P),has_meth(P,M,A,R),$\neg$dec_meth(C,M).
\end{lstlisting}

\end{document}